\documentclass[a4paper]{jpconf}%
\usepackage{graphicx,color}
\usepackage{amsmath,amssymb}
\usepackage[figuresright]{rotating}
\usepackage{amsmath}
\usepackage{amsfonts}
\usepackage{amssymb}
\usepackage{graphicx}%
\usepackage{subfigure}

\newcommand{\Sum}{\mathop{\Sigma}}
\begin{document}
\title{DRA method: Powerful tool for the calculation of the loop integrals.}
\author{R N Lee }
\address{The Budker Institute of Nuclear Physics\\
e-mail: \mailto{r.n.lee@inp.nsk.su}
}

\begin{abstract}
We review the method of the calculation of multiloop integrals
suggested in Ref.\cite{Lee2010}.
\end{abstract}

\section{Introduction}

The calculation of the multiloop integrals is important for the applications of the perturbative quantum field theory. A number of the calculational methods has been developed for this purpose. They roughly fall in to two categories: direct and indirect methods. The former include Feynman parameterization technique, Mellin-Barnes representation method, Gegenbauer polynomials expansion in coordinate space. The latter methods make use of the equations which are satisfied by the integrals. The construction of these equations relies on the possibility to make the IBP reduction of the integrals under consideration to some finite set of the integrals, called ``master integrals''. Using this technique one can derive the differential equation for the integrals with several scales, or difference equation for the integrals with one scale. These equations determine the integrals up to the solution of their homogeneous parts.

In Ref. \cite{Lee2010} a method of multiloop integrals evaluation
based on $\mathcal{D}$ recurrence relations \cite{Tarasov1996} and
$\mathcal{D}$-analyticity was suggested (DRA method). Since then it was applied to the calculation of various complicated integrals  \cite{Lee:2010cga,Lee:2010ug,Lee:2010ik,Lee:2010hs,Lee:2011jf,Lee:2011jt}.
In this contribution we review this method.

\section{DRA method}

\subsection{Dimensional recurrence relation}
Assume that we are interested in the calculation of the $L$-loop integral
depending on $E$ linearly independent external momenta $p_{1},\ldots,p_{E}$.
There are $N=L(L+1)/2+LE$ scalar products depending on the loop momenta
$l_{i}$:%

\begin{equation}
s_{ij}=s_{ji}=l_{i}\cdot q_{j}\,;\quad i=1,\ldots,L;\quad j=1,\ldots,K,
\end{equation}
where $q_{1,\ldots,L}=l_{1,\ldots,L}$, $q_{L+1,\ldots,K}=p_{1,\ldots,E}$, and
$K=L+E$.

The loop integral has the form%
\begin{align}
J^{\left(  \mathcal{D}\right)  }\left(n_1,\ldots,n_N\right)  =  &   \int\frac{d^{\mathcal{D}}l_{L}\ldots d^{\mathcal{D}}l_{1}}%
{\pi^{L\mathcal{D}/2}D_{1}^{n_{1}}D_{2}^{n_{2}}\ldots D_{N}^{n_{N}}}
\label{eq:J}%
\end{align}
where the scalar functions $D_{\alpha}$ are linear polynomials with respect to
$s_{ij}$. The functions $D_{\alpha}$ are assumed to be linearly independent
and to form a complete basis in the sense that any non-zero linear combination
of them depends on the loop momenta, and any $s_{ij}$ can be expressed in
terms of $D_{\alpha}$.

The integral $J^{\left(  \mathcal{D}\right)  }\left(n_1,\ldots,n_N\right) =J^{\left(  \mathcal{D}\right)  }\left(  \mathbf{n}\right)  $ can be considered as a value of  function of integer $N$-dimensional vector $\mathbf{n}=\left(n_1,\ldots,n_N\right)$. The set of points in $\mathbb{Z}^N$ (and the corresponding set of integrals), having the same set of positive coordinates, form a \textit{sector}. Thus the whole space is split into $2^N$ different sectors which can be conveniently labeled by their simplest elements. E.g., $(1,1,0,0,\ldots,0)$ denotes the sector in which all points have exactly two first coordinates positive. The points (and the corresponding integrals) in $\mathbb{Z}^N$ and the sectors form a partially ordered set with respect to the following relation. We say that the point $\mathbf{n}$ is \textit{simpler} than the point $\mathbf{m}$ iff all coordinates, positive in $\mathbf{n}$, are also positive in $\mathbf{m}$, but not vice versa.

It is convenient \cite{Lee2010} to introduce the operators $A_{i},B_{i}$ which act on such
functions as
\begin{align}
(A_{i}f)(\ldots,n_{i},\ldots)  &  =n_{i}f(\ldots,n_{i}+1,\ldots)\,,\nonumber\\
(B_{i}f)(\ldots,n_{i},\ldots)  &  =f(\ldots,n_{i}-1,\ldots)\,.
\end{align}
These operators respect the notion of sectors in a sense that the values of functions $A_{i}f$ and $B_{i}f$ in the point of some sector are expressed via the values of function $f$ in the same or simpler sectors. Using these operators, we can express the integral in $\mathcal{D}\pm 2$ dimensions via the integrals in $\mathcal{D}$ dimensions \cite{Lee2010LL}:
\begin{align}
J^{\left(  \mathcal{D}+2\right)  }\left(  \mathbf{n}\right)
&=\frac{(2\mu
)^{L}\left[ \det\left\{p_i\cdot p_j\right\}_{i,j=1\ldots E} \right]  ^{-1}}{\left(
\mathcal{D}-E-L+1\right)  _{L}} P\left(  B_{1},\ldots,B_{N}\right)  J^{\left(  \mathcal{D}%
\right)  }\left(  \mathbf{n}\right)\, ,\label{eq:Lowering}\\
J^{\left(  \mathcal{D}-2\right)  }\left(  \mathbf{n}\right)
&=(\mu/2)^{L}Q\left(  A_{1},\ldots,A_{N}\right)J^{\left(  \mathcal{D}%
\right)  }  \left(  \mathbf{n}\right)\, ,\label{eq:Raising}%
\end{align}
where $\mu=\pm1$ for the Euclidean/pseudoEuclidean case, and polynomials $P$ and $Q$ are determined as
\begin{align}
P(D_1,\ldots,D_N)&=
\det\left\{q_i\cdot q_j\right\}_{i,j=1\ldots K}\,,\nonumber\\
Q\left(\frac{\partial}{\partial D_1},\ldots,\frac{\partial}{\partial D_N}\right)
&=
\det\left\{  2^{\delta_{ij}}\frac{\partial}{\partial s_{ij}}\right\}_{i,j=1\ldots L}
\,.
\end{align}
These relations are the most useful when the integral in the left-hand side is master. Then, making the IBP reduction of the right-hand side of Eqs. \eqref{eq:Lowering},\eqref{eq:Raising}, one obtains difference equations for the master integrals. After this, we obtain a linear combination of the integrals of the same, or simpler, sectors as the integral in the left-hand side of Eqs. \eqref{eq:Lowering},\eqref{eq:Raising}. We assume that the integrals of the the simpler sectors are already known at this stage, either by the same, or by some other method.

The IBP reduction makes Eqs. \eqref{eq:Lowering},\eqref{eq:Raising} equivalent, so in what follows we consider only the lowering dimensional recurrence relation, obtained from Eq. \eqref{eq:Lowering}. We also change notations, writing
\begin{equation}
J^{(\mathcal{D})}=J(\nu),
\end{equation}
where $\nu=\mathcal{D}/2$. If a sector of the integral in the left-hand side contains only one master integral, the general form of the equation is the following:
\begin{equation}
J(\nu+1)=C(\nu)J(\nu)+R(\nu),
\end{equation}
where $C(\nu)$ is some rational function of $\nu$ and $R(\nu)$ is a
non-homogeneous part constructed of the simpler master integrals in $\mathcal{D}=2\nu$ dimensions.
If a sector contains several master integrals, it is convenient to consider them as a column-vector integral $\mathbf{J}(\nu)$ and to write the equations for them in a matrix form:
\begin{equation}
\mathbf{J}(\nu+1)=\mathbb{C}(\nu)\mathbf{J}(\nu)+\mathbf{R}(\nu),\label{eq:matrixDRR}
\end{equation}
where $\mathbb{C}(\nu)$ is some rational matrix (i.e., the matrix with coefficients being rational functions of $\nu$).

\subsection{General solution}
In order to write the general solution of this equation, we introduce the notion of indefinite sum and indefinite product.
Given the function $F(\nu)$, the \textit{indefinite sum} $\Sum F (\nu)$ is any function, satisfying the relation
\begin{equation}
\Sum F (\nu+1)-\Sum F (\nu)=F (\nu)
\end{equation}
The indefinite sum $\Sum F (\nu)$ is defined up to the addition of arbitrary periodic function of $\nu$.
If the function $F(\nu)$ decreases faster than $1/\nu$ when $\nu\to +\infty$  and/or $\nu\to -\infty$, one can write
\begin{equation}
\Sum F (\nu)=\Sum\limits_{+\infty} F (\nu)+\omega(z)\stackrel{\text{def}}{=}-\sum_{n=0}^{+\infty}F(\nu+n)+\omega(z)\label{eq:SumUp}
\end{equation}
and/or
\begin{equation}
\Sum F (\nu)=\Sum\limits_{-\infty} F (\nu)+\omega(z)\stackrel{\text{def}}{=}\sum_{n=-\infty}^{-1}F(\nu+n)+\omega(z)\label{eq:SumDown},
\end{equation}
where $\omega(z)=\omega(\exp(2\pi i \nu))$ is arbitrary periodic function.

The \textit{indefinite product} $\Pi F (\nu)$ is any function, satisfying the relation
\begin{equation}
\Pi F (\nu+1)/\Pi F (\nu)=F (\nu)
\end{equation}
$\Pi F (\nu)$ is defined up to the multiplication by arbitrary periodic function of $\nu$. Note that for any rational function $F(\nu)$ one can express $\Pi F(\nu)$ as a product of $\Gamma$-functions multiplied by arbitrary periodic function, see Ref. \cite{Lee2010}. We will determine the indefinite product  $\Pi \mathbb{F}(\nu)$  also for the matrix functions $\mathbb{F}(\nu)$ as a matrix function, satisfying the relation
\begin{equation}
\Pi \mathbb{F} (\nu+1)(\Pi \mathbb{F} (\nu))^{-1}=\mathbb{F}(\nu)
\end{equation}
Obviously, this defines $\Pi \mathbb{F} (\nu)$ up to the multiplication from the right by arbitrary periodic matrix.

In these notations one can write down the general solution of Eq. \eqref{eq:matrixDRR} in the form
\begin{equation}
\mathbf{J}(\nu)=\mathbb{S}(\nu)^{-1}\Sum \mathbb{S}(\nu)\mathbf{R}(\nu),\label{eq:generalSolution}
\end{equation}
where $\mathbb{S}(\nu)=[\Pi \mathbb{C}(\nu)]^{-1}$ is called a \textit{summing factor}. If the summand in Eq.\eqref{eq:generalSolution}
decreases faster than $1/\nu$ when $\nu\to +\infty$  and/or $\nu\to -\infty$, one can use Eqs. \eqref{eq:SumUp},\eqref{eq:SumDown} to write
\begin{equation}
\mathbf{J}(\nu)=\mathbb{S}(\nu)^{-1}\boldsymbol{\omega}(z)+\mathbb{S}(\nu)^{-1}\Sum_{\pm\infty} \mathbb{S}(\nu)\mathbf{R}(\nu)\,.\label{eq:generalSolution}
\end{equation}
Note that the second term in the right-hand side is invariant with respect to the choice of the summing factor. Thus, in order to find the integral, we need to fix the function $\boldsymbol{\omega}(z)$ for any choice of the summing factor. Here we note the essential difference between the general solutions of \emph{differential} and \emph{difference} equations. The homogeneous part of the general solution of the differential equation is parameterized  by finite number of constants, while the homogeneous part of the general solution of the difference equation is parameterized by the periodic function(s). While fixing the finite number of constants is a relatively simple task, fixing the periodic function may be quite difficult. The key point of the DRA approach is the usage of the analytical properties of the integral in order to fix this function. Let us rewrite Eq.\eqref{eq:generalSolution} as
\begin{equation}
\boldsymbol{\omega}(z)=\mathbb{S}(\nu)\mathbf{J}(\nu)-\Sum_{\pm\infty} \mathbb{S}(\nu)\mathbf{R}(\nu)\,.\label{eq:omega}
\end{equation}
Suppose that we can show that the right-hand side of this equation is a meromorphic  function of $z$ in the Riemann sphere. Then $\boldsymbol{\omega}(z)$ is determined, up to a constant,  by the principal parts of its Laurent series around singular points.
The analysis of the analytical properties of the right-hand side of Eq. \eqref{eq:omega} can be performed in any vertical stripe of unit length in the complex plane of $\nu$, as demonstrated in Fig. 1. There is no cut on the ray $\arg (z)=2\pi\nu_0$ because the right-hand side is necessarily a periodic function of $\nu$. The proper choice of this stripe (\textit{basic stripe}) can simplify the evaluation of the integral.
\begin{figure}
  \setlength{\unitlength}{1bp}%
  \centering
  \begin{picture}(340.60, 149.59)(0,0)
  \put(0,0){\includegraphics{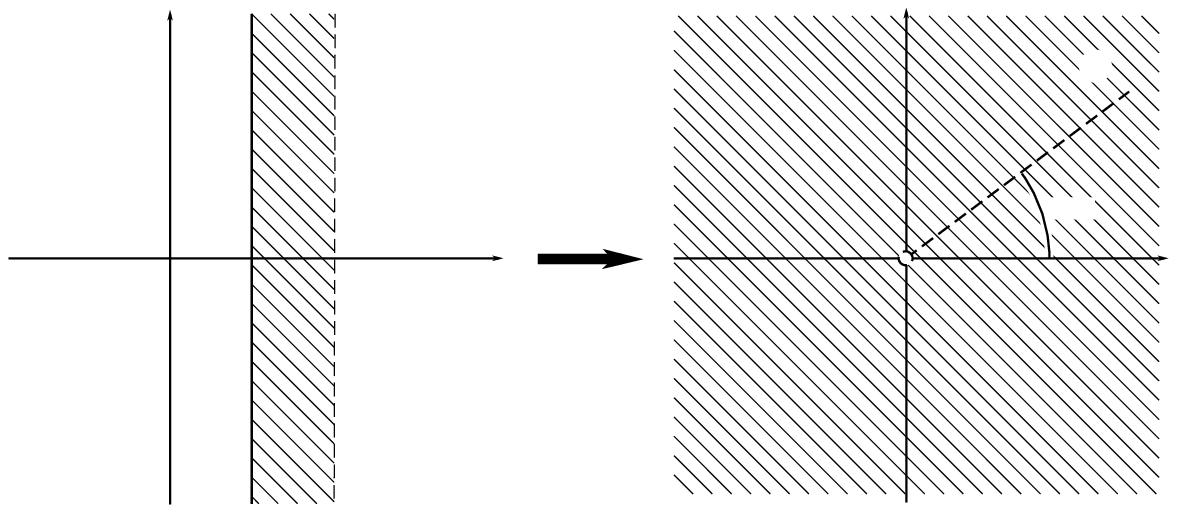}}
  \put(316.74,128.61){\fontsize{6.91}{8.30}\selectfont \makebox[0pt]{$z$}}
  \put(121.56,134.57){\fontsize{6.91}{8.30}\selectfont $\nu$}
  \put(73.17,77.42){\fontsize{3.71}{4.45}\selectfont \makebox[0pt][r]{$\nu_0$}}
  \put(98.70,77.36){\fontsize{3.71}{4.45}\selectfont $\nu_0+1$}
  \put(302.89,87.93){\fontsize{3.71}{4.45}\selectfont $2\pi\nu_0$}
  \end{picture}%
  \caption{Mapping $\nu\to z$}
\end{figure}

\subsection{Analytical properties of integrals}

Usually, one is interested in the analytical properties of the integral as a function of some external invariant. Here we are interested instead in the analytical properties of the integral as a function of $\nu=\mathcal{D}/2$. It appears that the latter properties are much simpler than the former.

Let us consider the parametric representation of some $L$-loop integral in
Euclidean space with $I$ internal lines (see, e.g., Ref. \cite{Itzykson1980}):%
\begin{equation}
J\left(  \nu\right)  =\Gamma\left(  I-L\nu\right)  \int
dx_{1}\ldots dx_{I}\delta\left(  1-{\textstyle\sum}x_{i}\right)
\frac{\left[  q\left(  x\right)  \right]  ^{\nu L-I}%
}{\left[  p\left(  x\right)  \right]  ^{\nu\left(  L+1\right)-I}%
}.\label{eq:parametric}%
\end{equation}
The polynomials $q\left(  x\right)  $ and $p\left(  x\right)  $ are determined
by the graph. For our consideration it is important only that both these
functions are nonnegative in the whole integration region,%
\[
q\left(  x\right)  \geqslant0,\qquad p\left(  x\right)  \geqslant0.
\]
Suppose that the parametric representation converges for all $\nu$
in some interval $\left(\nu_{1},\nu_{2}\right)$.
Then it is easy to see that it converges on the whole stripe $S=\left\{  \nu
,\quad\operatorname{Re}\nu\in\left(\nu_{1},\nu_{2}\right) \right\}$. Indeed, we can estimate
\begin{equation}
\left\vert J\left(  \nu\right) \right\vert  \leqslant
\frac{\left\vert \Gamma\left(  I-L\nu\right)  \right\vert
}{\Gamma\left(  I-L\operatorname{Re}\nu\right)  }J\left(  \operatorname{Re}\nu\right)\,.
\label{eq:bound}
\end{equation}
Therefore, Eq. \eqref{eq:parametric} determines $J\left(\nu\right)$ as a
holomorphic function on the whole stripe $S$. In fact, in Ref. \cite{Gelfand1969} a more general fact was proved: the representation \eqref{eq:parametric} determines the meromorphic function on the whole complex $\nu$-plane.
The determination of poles position can be automatized thanks to \texttt{FIESTA} \cite{SmiSmTe2009}.

As $\operatorname{Im}\nu$ tends to $\pm\infty$ while
$\operatorname{Re}\nu$ is
fixed, Eq. \eqref{eq:bound} shows that
\begin{equation}
\left\vert J\left(  \nu\right) \right\vert
\lesssim\mathrm{const}%
\times e^{-\frac{\pi L\left\vert \operatorname{Im}\nu\right\vert }{2}}
\left\vert \operatorname{Im}\nu\right\vert ^{I-1/2-L\operatorname{Re}\nu}
<\mathrm{const}\times \left\vert z \right\vert^{\mp (L+0)/4}\,.
\end{equation}
Thus, the parametric representation can be used for the determination of the analytical properties of the first term in the right-hand side of Eq. \eqref{eq:omega}. Since the second term in the right-hand side of Eq. \eqref{eq:omega} is constructed of the integrals already known at this stage, we, indeed, can determine the position of the poles of function $\omega(z)$.

\subsection{Numerical issues}

\begin{figure}
\centering
\includegraphics[width=0.8\textwidth]{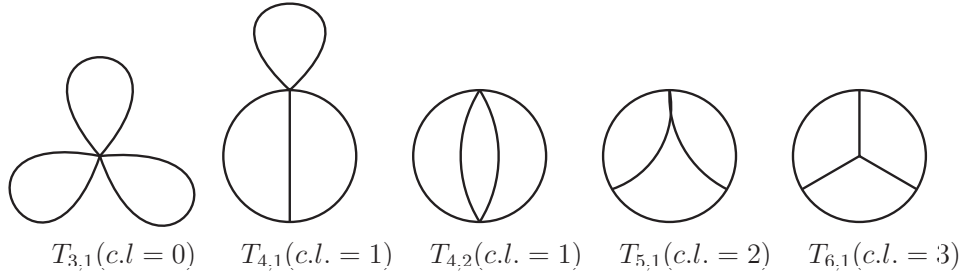}
  \caption{Three-loop all-massive tadpole master ntegrals.}
  \label{fig:masters}
\end{figure}

In order to demonstrate how the representations obtained within the DRA method can be treated numerically, let us consider the three-loop master integrals depicted in Fig. \ref{fig:masters}. Direct application of the DRA method gives the following results:
\begin{align}
T_{3,1}&=\Gamma (1-\nu )^3\label{eq:t31}\\
T_{4,1}&=\frac1{S_{4,1}}\left\{
\frac{4 \pi }{3 \sqrt{3}}-\Sum_{-\infty}S_{4,1}T_{3,1}
\right\}\,,
\quad S_{4,1}=\frac{3^{-\nu }}{\Gamma (2-2 \nu ) \Gamma (1-\nu )}\\
T_{4,2}&=\frac1{S_{4,2}}\left\{
\frac{3 \pi ^2}{16}
-
\Sum_{-\infty}\frac{(11 \nu-8) S_{4,2}T_{3,1}}{8 (\nu-1)}
\right\}\,,
\quad S_{4,2}=\frac{4^{1-3 \nu } \Gamma (2-\nu )^2}{\Gamma (3-3 \nu ) \Gamma (3-2 \nu )}
\\
T_{5,1}&=\frac1{S_{5,1}}\left\{
\frac{16 \pi ^2}{27}
+\Sum_{-\infty}S_{5,1}
\frac{4 (7 \nu-4 ) T_{3,1}-36 (3 \nu-2 ) T_{4,1}-(2+13\nu ) T_{4,2}}{27 (3 \nu-2 )}
\right\}\,,\nonumber\\
&{\,}\hspace{8cm} S_{5,1}(\nu)=\frac{9^{-\nu } \Gamma (1-\nu )}{\Gamma (2-2 \nu )^2}\\
T_{6,1}&=\frac1{S_{6,1}}\left\{
2 \pi ^2
+\Sum_{-\infty}\frac{S_{6,1}}{2} (T_{4,2}-3 T_{5,1})
\right\}\,,\quad S_{6,1}=\frac{2^{-\nu }}{\Gamma (2-2 \nu ) \Gamma (2-\nu )}\label{eq:t61}
\end{align}

The above results for the integrals $T_{5,1}$ and  $T_{6,1}$ contain repeated sums of the form:
\begin{align}
s=\sum_{n_1=0}^{\infty}\sum_{n_2=n_1}^{\infty}\ldots \sum_{n_k=n_{k-1}}^{\infty}
f_1(n_1)f_2(n_2)\ldots f_k(n_k)
\end{align}

\begin{figure}
\subfigure[~]{\includegraphics[width=5cm]{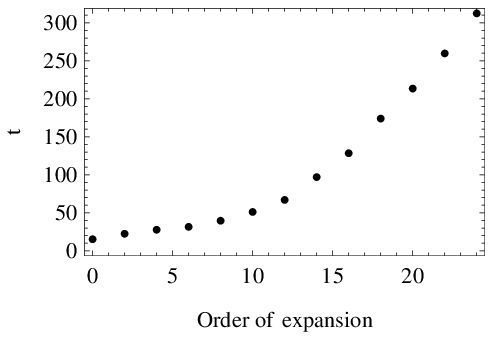}}
\subfigure[~]{\includegraphics[width=5cm]{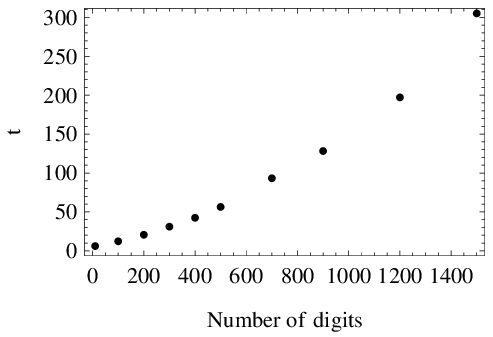}}
\subfigure[~]{\includegraphics[width=5cm]{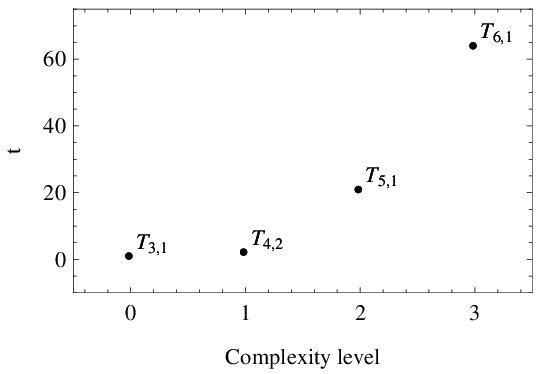}}
  \caption{Time of calculation (sec) as a function of
  (a) requested $\epsilon$-order ($T_{6,1}(4-2\epsilon)$  with 200-digit precision);
  (b) requested precision ($T_{6,1}(4-2\epsilon)$  up to $O(\epsilon^3)$);
  (c) complexity level (expansion up to $O(\epsilon^4)$ with 500-digits precision)}
  \label{fig:timing}
\end{figure}

From the viewpoint of numerical calculation, it is important that the dependence on the summation variables is factorized in the summand.
We note that it is not the case for the Mellin-Barnes multiple sums. Let us show that the calculation of such factorized sums can be performed without nested loops. Indeed, let us define $k$ sequences $s_1(n),\ldots,s_k(n)$ as follows:
\begin{align}
s_i(-1)=0\,,\quad
s_1(n\geqslant0)=s_1(n-1)+f_1(n-1)\,,\quad s_{i>1}(n\geqslant0)=s_{i}(n-1)+f_i(n)s_{i-1}(n)
\end{align}
It is easy to check that the last sequence, $s_k(n)$, converges to $s$. Obviously, the iterative calculation of these quantities can be organized in one loop. Therefore, the complexity of the numerical calculation grows moderately with the complexity level of the integrals. The dependence of the calculation time on the order of the expansion, required precision, and complexity level of the integrals is shown in Fig. \ref{fig:timing}. We see that for all three cases the method shows quite moderate growth of the evaluation time.

\section{Conclusion}

We have briefly reviewed the method of calculation of multiloop integrals
based on the $\mathcal{D}$-recurrence and $\mathcal{D}$-analyticity. The
method appears to be powerful enough to deal with the most complicated cases.

\ack
This work was supported by RFBR (grants Nos.  10-02-01238, 11-02-01196, 11-02-00220, 11-02-08379)
and by Federal special-purpose program ``Scientific and scientific-pedagogical personnel of innovative Russia''.
I appreciate the organizers support  for the participation in the ACAT conference.
I am grateful to Y. Schr\"oder for drawing my attention to typos in the previous version of this paper.

\end{document}